# Natural movement with concurrent brain-computer interface control induces persistent dissociation of neural activity

Luke BASHFORD[1,2,*], Jing WU[3], Devapratim SARMA[3], Kelly COLLINS[4], Jeff OJEMANN[4], Carsten MEHRING[2]

[1]Imperial College London, Bioengineering, UK; [2]Bernstein Centre & Faculty of Biology & BrainLinks-BrainTools, Univ. of Freiburg, Germany; [3]Bioengineering, Ctr. for Sensorimotor Neural Eng., [4]Dept. of Neurolog. Surgery, Ctr. for Sensorimotor Neural Eng., Univ. of Washington, USA.
*Corresponding author LB: luke.bashford11@imperial.ac.uk

## Introduction

As Brain-computer interface (BCI) technology develops it is likely it may be incorporated into protocols that complement and supplement existing movements of the user [1]. Two possible scenarios for such a control could be: the increasing interest to control artificial supernumerary prosthetics [2,3], or in cases following brain injury where BCI can be incorporated alongside residual movements to recover ability [4]. In this study we explore the extent to which the human motor cortex is able to concurrently control movements via a BCI and overtly executed movements. Crucially both movement types are driven from the same cortical site. With this we aim to dissociate the activity at this cortical site from the movements being made and instead allow the representation and control for the BCI to develop alongside motor cortex activity. We investigated both BCI performance and its effect on the movement evoked potentials originally associated with overt execution.

## Methods

Patient's undergoing epilepsy monitoring with subdural Electrocorticography (ECoG) signed informed consent to participate in the studies. All experiments were approved by the ethics committees at the University of Washington and Seattle Children's Hospital. The experiment contained three parts; preceding the BCI task all subjects performed a 'pre' movement screening. Subjects then performed either a '1D BCI' task and/or the 'Concurrent BCI' task, followed by a 'post' movement screening identical to the 'pre'. We present data from 4 subjects; 1 who performed concurrent only, 1 who performed 1D only, and 2 who performed both on separate days.

*Pre/Post Screening*: Subjects were given a visual cue to perform a repeated movement. The first cohort of subjects performed a force matching task, squeezing a force sensor to match a predefined value. The second cohort of subjects performed a self-paced tapping of the index finger. In each case subjects were cued to make the movement over 6s, repeated 20 times with a rest period of no movement lasting 2-4s (randomly selected to avoid anticipation) in between each trial. This task was performed identically before and after each BCI session to measure a baseline for movement only evoked activity, allowing us to determine changes caused by performing the BCI task. In an initial screening this task was used to select the single channel to control the BCI. The channel selected was that which had the largest 70-90Hz amplitude. The position of the electrode was also confirmed to be over motor cortex using MRI reconstructions of electrode locations where possible.

*1D BCI*: Brain activity was recorded using a TDT RZ5D processor and a PZ5 ADC (Tucker-Davis Technologies, Alachua, Florida, USA), or g.USBAmps (g.tec medical engineering GmbH, Schiedlberg, Austria) sampled at 1220 Hz and 1200 Hz, respectively. The vertical cursor movement was controlled by the normalized output of an auto-regressive filter using the spectral power in the 70-90Hz band of the recorded signal. Two targets were visible at the top and the bottom of a workspace. Subjects were instructed that movement imagery, with no overt execution, would move the cursor upward, while rest would move the cursor downward.

*Concurrent BCI*: Subjects were required to move a cursor from the centre of the workspace to one of 8 targets presented around a circle. The vertical component of cursor movement was controlled via the BCI (identical to the 1D task above). The horizontal component of the movement was controlled by repeatedly pressing a computer keyboard key with the index finger contralateral to the BCI control electrode location. Continuous tapping was required otherwise the cursor would automatically move back towards the centre of the workspace at a constant rate. The cursor position was updated at 30Hz, at each update the cursor position moved *(key press distance * number of presses) – return distance*, where key press distance is 0.87% and return distance is 0.026% of the total workspace 0-1(AU). Subjects had to achieve a concurrent control to reach the off vertical targets. Subjects completed 48 trials, of equal presentation of the 8 targets. This run was repeated as many times as possible per session. Subjects had 10s to complete each trial before a time out, followed by a 4s inter trial interval.

## Results

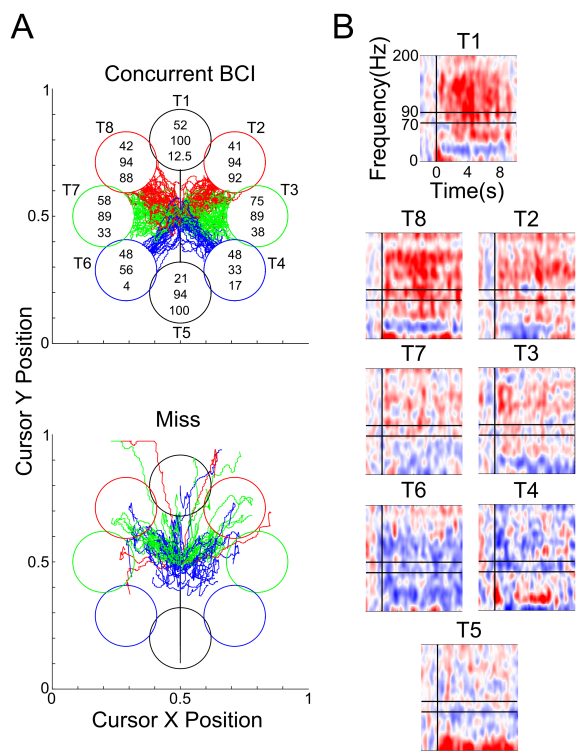

Fig1. A) Top; Cursor trajectories and individual subject target accuracy (row per subject) for 3 subjects during concurrent BCI. Bottom; Example 'miss' trajectories. B) STFT averaged over all successful trials of an exemplary session for one subject.

Subjects were able to gain control using our concurrent BCI task (Fig1A top). As the control site used was also activated by hand movement execution, an increased gamma power produced by the movement execution should drive the cursor upwards; this is evident in the example 'miss' trajectories (Fig1A bottom). In the successful trials therefore dissociation of the activity at the control electrode site from finger movements was required to move

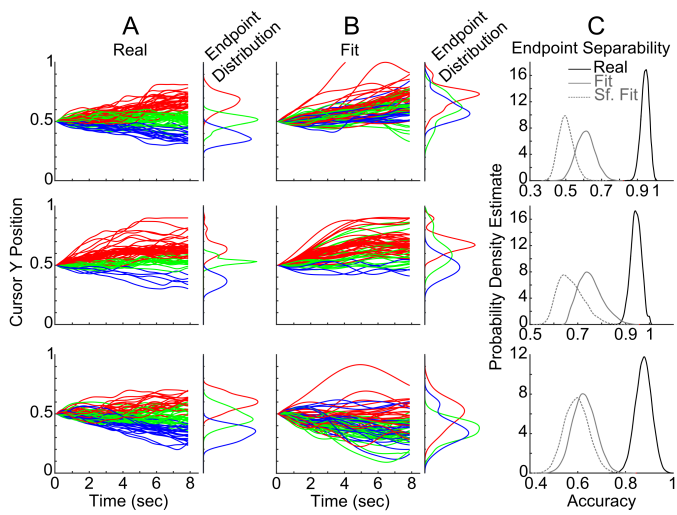

*Fig2. Rows show 3 subjects (Sub1 – first row, etc.). Column A) Smoothed real cursor Y position with endpoint distribution. B) Modelled cursor Y position with endpoint distribution. C) Accuracy of endpoint separability based on fitting 2 thresholds.*

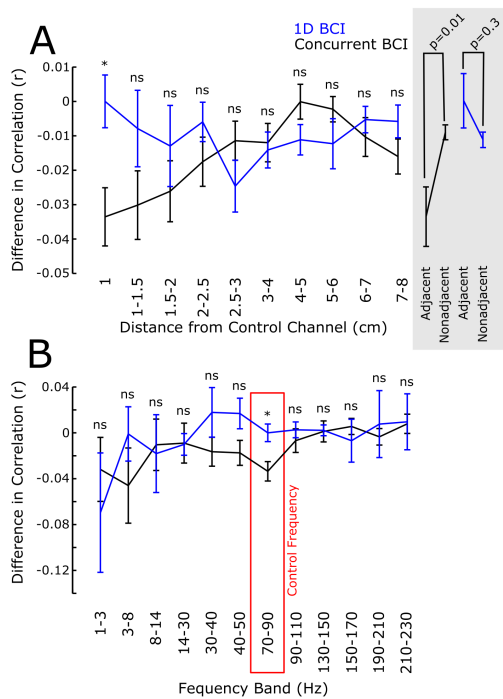

*Fig3. A) Correlation change between 'pre' and 'post' movement screening at the control frequency for electrodes at different distances from the control channel. Inlay shows pooled data within the group. B) Correlation change for concurrent and 1D BCI at adjacent electrodes across different frequencies. Error bars show SEM.*

the cursor in both vertical directions independently of the requirement to tap the finger for horizontal control (Fig1B). To validate that the movement execution and BCI control occurred independently, the Y position of the cursor (Fig2A), under BCI control, should not be influenced by movements. We made a best fit linear model between the change in Y position around the key presses and used this to recreate the cursor Y trajectories from the key presses (Fig2B), the recreated trajectories were then scaled up to match the endpoint standard deviation of the real trajectories. If the control was not independent we would expect a high correlation between real and fit trajectories, however we found a correlation within the chance distribution produced from a 10000 fold shuffle fit per run (Sub1; R=0.49, 92.1%ile. Sub2; R=0.78, 98.2%ile. Sub3; R=0.48, 77.6%ile. Fit and percentile in shuffle distribution respectively). Furthermore the fitted trajectories were not able to capture the endpoint separability seen in the real data. We fitted two thresholds to the data that maximised endpoint accuracy classification over a 10000 sample bootstrapping. The distribution of accuracies produced showed significantly lower endpoint accuracies in the model fit trajectories vs the real (Sub1 p<0.0005, Sub2&3 p<0.01) (Fig2C).

We further hypothesised that the dissociation of the BCI control activity from the finger movement induces a persistent neural reorganisation. To investigate this hypothesis, we correlated the 70-90Hz (control frequency band) amplitude envelope of the neural signal between the control and other channels and computed the change of the correlation between 'pre' and 'post' screening. Our results showed a clear reduction in the correlation from 'pre' to 'post' between the control channel and adjacent channels (distance of 1cm away) compared to nonadjacent channels (distance >1cm) only for concurrent but not for 1D BCI sessions (pooled data p=0.01, p=0.3 respectively, Wilcoxon rank sum test, Holm-Bonferroni corrected) (Fig3A inlay). Comparing between the groups demonstrated a significant difference in correlation also at adjacent electrodes only (p=0.013) (Fig3A). This effect decreases with increasing distance. We further compared the signal correlation to adjacent channels across different frequency bands and show this effect was also unique to the control frequency band (p=0.013) (Fig3B). This suggests that it is not general BCI use but rather the concurrent control that induces a dissociation of neural activity specifically from the control site that persists beyond the concurrent control task.

**Discussion**

We demonstrated that human subjects are able to gain an independent and concurrent control of a BCI and overt movement execution. Furthermore we revealed that this concurrent control, unlike typical BCI only paradigms, enforces dissociation between the neural control signal and the overt movement evoked neural activity. Moreover, we showed that this dissociation of neural activity persists beyond the concurrent control task. We therefore propose that it reflects a change in the cortical organisation indicating that a distinct mapping or representation of BCI control can develop amongst concurrently active cortical areas. This provides a novel extension of previous ideas of map formation in BCI use [5]. The potential for concurrently controlled BCI and movement execution has until now only been demonstrated in primates [6,7], we extend this to the human case and give a novel demonstration of the physiological changes at the control site due to concurrent control. This framework demonstrates the potential for the control of BCIs in addition to natural movements, for example a future 'third arm' BCI control.

**Acknowledgements:** This work was funded by Bernstein Focus Neurotechnology Fr/Tu, BrainLinks-BrainTools Cluster of Excellence and Center for Sensorimotor Neural Engineering, University of Washington.